%% file: root.tex
\documentclass[letterpaper, 10 pt, conference]{ieeeconf}  % Comment this line out if you need a4paper

\IEEEoverridecommandlockouts                              % This command is only needed if 
\usepackage{amsmath}
\usepackage{amssymb}

\usepackage{amsthm}
\usepackage[thinc]{esdiff}
\usepackage{subcaption}
\usepackage{cleveref}
\usepackage{mathtools}
\usepackage{graphicx}
\usepackage[dvipsnames]{xcolor}
\definecolor{review}{RGB}{0,0,255}

\newcommand\R{\mathbb{R}}

\newtheoremstyle{style}%                % Name
  {}%                                     % Space above
  {}%                                     % Space below
  {\itshape}%                             % Body font
  {}%                                     % Indent amount
  {\bfseries}%                            % Theorem head font
  {.}%                                    % Punctuation after theorem head
  { }%                                    % Space after theorem head, ' ', or \newline
  {}%                                     % Theorem head spec (can be left empty, meaning `normal')
\theoremstyle{style}

\newtheorem{remark}{Remark}

\usepackage{algorithm}
\usepackage{algpseudocode}

\input{defs}

\title{\LARGE \bf
The Minkowski Wrap: A Relativistic Speed Limiter
}

\author{Nicoletta Prencipe$^1$, Ba\c{s}ak Sak\c{c}ak$^{2}$ and Steven M. LaValle$^3$ % <-this % stops a space
\thanks{This work was supported by a European Research Council Advanced Grant (ERC AdG, ILLUSIVE: Foundations of Perception Engineering, 101020977), Academy of Finland (project BANG! 363637).}%, CHiMP 342556).}% <-this % stops a space%
\thanks{$^{1}$CNRS, Institut de Recherche Mathématique de Rennes, University of Rennes, {\tt\small nicoletta.prencipe@univ-rennes.fr}}
\thanks{$^{2}$Department of Advanced Computing Sciences, Maastricht University, {\tt\small basak.sakcak@maastrichtuniversity.nl}}
\thanks{$^{3}$Faculty of Information Technology and Electrical Engineering, University of Oulu, %P.O. Box 4500, Oulu, FI-90014, Finland 
{\tt\small steven.lavalle@oulu.fi}}
}

\begin{document}

\maketitle
\thispagestyle{empty}
\pagestyle{empty}

%%%%%%%%%%%%%%%%%%%%%%%%%%%%%%%%%%%%%%%%%%%%%%%%%%%%%%%%%%%%%%%%%%%%%%%%%%%%%%%%
\begin{abstract}
In special relativity, a particle can experience constant acceleration, but at the same time, its motion is constrained by the velocity limit imposed by the speed of light $c$. 
Inspired by this principle, we propose a method for enforcing velocity bounds in control systems by replacing $c$ with the maximum attainable speed of the system. We refer to this as the ``Minkowski wrap," the operation of deforming the phase portrait of a system so as to enforce desired speed limits. We apply this idea to shape the input generated by a state-feedback stabilizing controller and time-optimal controller considering controlling a double integrator system.
By applying Pontryagin's Maximum Principle, we show that the time-optimal control of a wrapped double-integrator system is bang-bang. 
The proposed method transforms the classical double-integrator dynamics into a ``wrapped" system that respects velocity bounds without the need for clipping, offering an explicit nonlinear feedback control strategy conducive to safety applications. 
\end{abstract}
%%%%%%%%%%%%%%%%%%%%%%%%%%%%%%%%%%%%%%%%%%%%%%%%%%%%%%%%%%%%%%%%%%%%%%%%%%%%%%%%
\section{Introduction}
Simulations that depict how the world would appear if the speed of light $c$ were much smaller than it actually is have proven useful, particularly in science outreach, to visualize concepts from special and general relativity. Can this be more than just an outreach exercise? The second postulate of the theory of special relativity states that the speed of light in a vacuum is the same for all inertial observers. This leads to the fact that the set of possible speeds is bounded by $c$, i.e. the set of relativistic velocities is a ball of radius $c$.

Consider a control system with unbounded speed, to which we impose a bounded speed \( v_{\max} \), analogous to a mechanical \textit{governor} or \textit{speed limiter} (see e.g., \cite[Ch.~1]{Astrom:08}). The idea is to use constructs inspired by special relativity to introduce velocity bounds, replacing the speed of light \( c \) with an arbitrary maximum speed \( v_{\max} \), in such a way that the phase portrait of the bounded system is ``warped" through relativistic constructs instead of being clipped at \( v_{\max} \).

The mathematical frameworks underlying physical theories were developed in parallel with, and in some cases even independently of, their applications. Viewed in this way, they may be suitable for types of applications different from those for which they were originally conceived. In this sense, \( c \) can be seen as merely a parameter of the relativistic model.

This type of approach has been adopted, for instance, in the context of modeling human visual perception (such as color \cite{Yilmaz:62} or speed perception \cite{Caelli:78}) where only special relativity was needed. In those cases, the speed of light corresponds, respectively, to the maximum color saturation or to the maximum perceivable speed.
%This type of approach has been adopted, for instance, in the context of modeling human visual perception, such as color \cite{Yilmaz:62} or speed perception \cite{Caelli:78}. In those cases, the speed of light corresponds, respectively, to the maximum color saturation or to the maximum perceivable speed, which is clearly lower than the physical speed of light. For those only special relativity was needed. 

Considering optimal control under relativistic systems, 
Pontryagin’s Maximum Principle has been applied to rocket trajectory optimization under relativistic constraints with different cost functions and dynamics \cite{Henriques:12,Anderson:66}.

A recent line of research explored theoretical analogies between robot control theory and general relativity. 
Li et al.~\cite{Li:22, Li:23} showed that the classic metaphor of curved spacetime, often illustrated as a sphere deforming an elastic membrane, shares some formal similarities with the description of the dynamics of robots moving on non-flat, deformable surfaces. In this framework, a wheeled robot naturally follows geodesic paths, with the membrane's curvature depending on the robot’s velocity and the surface’s elasticity. Robots moving in a dynamic, non-flat environment with which they can interact and deform turn out to behave like “test particles in a fictitious spacetime”. Reinhardt et al.~\cite{Reinhardt:25} also propose the concept of ``artificial spacetimes,” as the adaptation of techniques from general relativity to the control of multiple resource-limited robots. In this framework, reactive robots moving within control fields exhibit trajectories whose dynamics correspond to those of light rays in general relativity. 

Imposing constraints on the trajectories of a control system is relevant for a broad range of practical applications with safety concerns, for example, control of energy systems~\cite{Magni01}, robot motion control~\cite{Wieber:06}, and autonomous driving~\cite{Ames:14}. A large percentage of the methods in the literature addressing constraints is based on Model Predictive Control~\cite{Rawlings:17}, in which a finite-dimensional constrained optimization problem is solved in a receding horizon manner. In this case, the control law is implicitly state-feedback. Considering explicit feedback, a more recent line of work is focused on augmenting Control Lyapunov Functions (CLFs) with Control Barrier Functions (CBFs) (\cite{Ames:19,Xiao:22}) to guarantee constraint satisfaction. In this case, the joint design of CLF-CBF is posed as an optimization problem. Finally, building upon an existing stabilizing controller, reference governors (see \cite{Garone:17} for a review) shape the reference to the controller to guarantee that the constraints are satisfied. 

In the present work, we consider a second order system with a desired bounded velocity and show how tool from special relativity suffice to impose this bound, or ``wrap'' the system with a safeguard.

In particular, Section \ref{sec:rel} introduces the relativistic version of the double integrator and its solutions, which follow directly from the relativistic second law of dynamics, also known as Minkowski law. The classic double integrator model, $\ddot{q} = u$, becomes problematic under relativistic constraints, since a particle subject to constant acceleration would eventually exceed the speed limit $c$. However, a particle can maintain constant \textit{proper} acceleration, that is, the acceleration experienced by the particle itself, while its motion, as described in an inertial reference frame, remains within the velocity bound. Section \ref{sec:pontryagin} demonstrates how Pontryagin's Maximum Principle applies to the relativistic double integrator and shows that a bang-bang time-optimal control can be obtained. Section \ref{sec:nonlin} is dedicated to showing how this relativity-inspired way of bounding the velocity can be used as a sort of \textit{feedback nonlinearization} technique for safety purposes, meaning that the control input is reshaped in a way that intentionally introduces nonlinear behavior in the dynamics to enforce the velocity bound. Finally, possible generalizations of this relativistic speed-wrapping methodology to more complex systems are proposed.

\section{The relativistic double integrator}\label{sec:rel}
In this section, we introduce some key notions from special relativity, with particular emphasis on the dynamics of a particle undergoing constant proper acceleration under relativistic constraints, and on the comparison of its behavior with that of the standard double integrator under constant controls.

In special relativity, the trajectory of a particle described in an inertial reference frame cannot have constant \textit{coordinate acceleration} $\ddot q$, since this would eventually violate the speed limit imposed by the theory. However, it is possible for a particle to experience a constant \textit{proper acceleration} $\alpha$, that is, the acceleration that would be indicated by an accelerometer co-moving with the particle.

The classical second law of dynamics, \( m\ddot q = F \), where \(F\) and \(m\) denote force and mass, respectively, no longer holds as \( |\dot q| \rightarrow c \). In special relativity, the corresponding formulation, known as the \textit{Minkowski equation} or law, is
\begin{equation}\label{eq:dyn}
    \frac{d}{dt}(m\gamma \dot q) = F,
\end{equation}
where the Lorentz factor $\gamma$ that corrects the velocity is
\begin{equation}\label{eq:gam}
    \gamma = \frac{1}{\sqrt{1 - (\dot q / c)^2}}.
\end{equation}
Notice that for $|\dot q| \ll c$, Eq.~\eqref{eq:dyn} becomes the classical Newton's law, as $\gamma\rightarrow 1$. Expanding the derivative in Eq.~\eqref{eq:dyn} yields, for the case of one-dimensional motion, the equivalent expression $ m\ddot q\gamma^3 = F$. See also \cite{Barone:04,Gourgoulhon:16,Gallant:12}.

A consequence of the relativistic form of the second law of dynamics is that, for a particle undergoing one-dimensional motion, the coordinate acceleration and the proper acceleration are related by the following expression
\begin{equation}\label{eq:alph}
    \ddot q \gamma^3 = \alpha
\end{equation}
with $\gamma$ as defined in Eq.~\eqref{eq:gam}.

In the following, we denote the constant proper acceleration by \(u\), which will serve as the control parameter, while retaining \(c\) for the speed of light, although our intent is to eventually replace it with the control system’s speed bound \(v_{\max}\). Eq.~\eqref{eq:alph} can then be reformulated as
\begin{equation}\label{eq:reldi}
    \begin{cases}
        \ddot q = u \left(1-\frac{\dot q^2}{c^2} \right)^{3/2}\\
        %\dot q(0), q(0) = \dot q_0, q_0
        \dot q(t_0), q(t_0) = \dot q_0, q_0
        %\dot q(t_0)= \dot q_0 \\
        %q(t_0) = q_0        
    \end{cases},
\end{equation}
where $t_0\geq 0$ and %$u:[t_0,t_f]\rightarrow U$, and 
$u\in U=[u_{\min},u_{\max}]$ is a real interval containing the origin.

In the rest of this paper, we will refer to the system whose dynamics is described by the previous differential equation as the \textit{relativistic double integrator}, since it represents the closest analogue to the classical double integrator, $\ddot q = u$, while fulfilling the relativistic constraints. In a sense, the particle ``experiences" a form of double-integrator dynamics, undergoing uniform proper acceleration.

Solving Eq.~\eqref{eq:reldi}, the velocity and position of the particle are given, respectively, by the following equations: 
\begin{equation}\label{eq:qdot}
    \dot q (t) = \frac{\dot q_0\gamma_0 + u(t-t_0)}{\mu(u,t)},
\end{equation}
\begin{equation}\label{eq:q}
    q(t)= q_0 + \frac{c^2}{u}\left(\mu(u,t) - \gamma_0\right),
\end{equation}
where the terms $\gamma_0$ and $\mu(u,t)$ are given by
\begin{equation}\label{eq:gammu}
    \gamma_0 = \frac{1}{\sqrt{1-\frac{\dot q_0^2}{c^2}}}, \quad \mu(u,t)=\sqrt{1+  \frac{(\dot q_0\gamma_0 + u(t-t_0))^2}{c^2}}.
\end{equation}
Eq.~\eqref{eq:q} is referred to as \textit{hyperbolic motion} and is attributed to Minkowski \cite{Minkowski:09,Rindler:91}.

\begin{remark}\label{rem:bounded_vel}
In Eq.~\eqref{eq:qdot}, notice that, for any $u\in U$, $|\dot{q}(t)| \rightarrow c$, as $t \rightarrow +\infty$, meaning that, under constant proper acceleration, it takes infinite time to reach the speed of light. The same holds as well for the case of non-constant proper acceleration $u:[t_0,t_f]\rightarrow U$.
\end{remark}

\begin{figure}%[htbp]
    \centering
    \includegraphics[width=1\columnwidth]{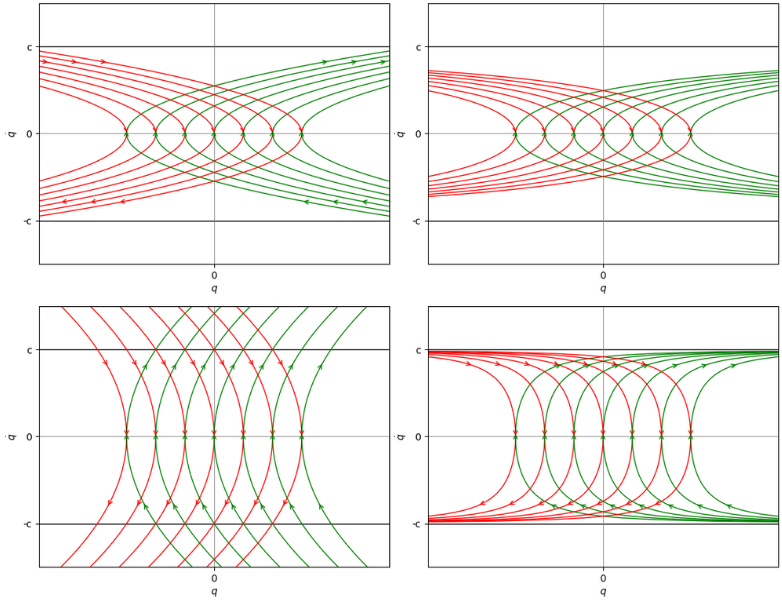}
    \caption{Comparison of the phase portraits of the standard and relativistic double integrators (first and second column, respectively). The red curves correspond to $u \equiv -kc$, and the green curves to $u \equiv kc$, with $k =0.5 ,\,5$ (first and second row, respectively). %0.5c,\, c,\, 2c,$ and $5c$.
    Note that for the relativistic double integrator the trajectories are confined to the region $|\dot{q}| < c$. }
    \label{fig:all_phases}
    \vspace{-2em}
\end{figure}

It is possible to express $q$ as a function of $\dot q$ obtaining
\begin{equation}\label{eq:phaseportr}
    q(\dot q) = q_0 + \frac{c^2}{u}\left( \left(1-\frac{\dot q^2}{c^2}\right)^{-1/2} -\gamma_0\right).
\end{equation}
Fig.~\ref{fig:all_phases} (second column) shows the phase portrait of the system when $u_\text{max} = -u_\text{min}$; the green curves are obtained by fixing $u \equiv u_\text{max}$, and the red curves by fixing $u \equiv u_\text{min}$.

The phase portrait of the classic double integrator is
\begin{equation}\label{eq:phaseportrDI}
    q(\dot q) = q_0 + \frac{\dot q_0}{u}(\dot q - \dot q_0) + \frac{1}{2u}(\dot q - \dot q_0)^2.
\end{equation}

It is well known that for a fixed, constant $u$, the phase portrait of the double integrator consists of parabolas (see the first column in Fig.~\ref{fig:all_phases}), directed toward the positive $q$ axis when $u$ is positive, and toward the negative axis when $u$ is negative.

If we take the same constant controls for both the standard and relativistic double integrators, and compare their phase portraits, as depicted in Fig.~\ref{fig:all_phases}, we can see that when $|u| \ll c$, the classical and relativistic double integrators look alike, see the case $|u| = 0.5c$. As $|u|/c$ increases, the parabolas of the standard double integrator increasingly violate the speed bound, while the trajectories of the relativistic one become more and more compressed toward the velocity limit (see the second column in Fig.~\ref{fig:all_phases}).

\section{Time-optimal control for the relativistic double integrator}\label{sec:pontryagin}
In this section, we apply Pontryagin's Maximum Principle (PMP), see e.g.~\cite{Liberzon:12}, to characterize the time-optimal, piecewise-constant (bang-bang) control for the relativistic double integrator.

For simplicity, we have so far treated \(u\) as a constant parameter. 
We now consider \(u\) as a time-dependent control function 
\(u:[t_0,t_f]\rightarrow U\), where the admissible control set \(U\) 
is the interval \(U=[u_{\min},u_{\max}]\subset\mathbb{R}\) with \(u_{\min}<0<u_{\max}\). 

Let \(x_1 = q\) and \(x_2 = \dot{x}_1 = \dot{q}\). 
Then, the Hamiltonian of the double integrator is given by
\begin{equation}\label{eq:hamDI}
    H(x,u,p,p_0) = p_0 + p_1 x_2 + p_2 u,
\end{equation}
where $p_0\leq 0$ is a real constant, $p=(p_1,p_2)$ with $p:[t_0,t_f]\rightarrow \mathbb{R}^2$ being the costates.

It is well known, from the bang-bang principle for linear systems and as a direct consequence of the PMP, that, for a fixed-endpoint control problem, the time-optimal control of the double integrator is of bang-bang type; that is, it is piecewise constant (taking values only at the extrema of the control interval) with at most one switch.

In accordance with the PMP, if an optimal control \(u^*:[t_0,t_f]\rightarrow U\) exists, it maximizes the Hamiltonian pointwise in time. This means that for each \(t\in[t_0,t_f]\), the function 
\(u \mapsto H(x^*(t),u,p^*(t),p_0^*)\) attains its global maximum at \(u=u^*(t)\). 
Thus, for all \(u \in U\), and for all \(t \in [t_0,t_f]\),
\begin{equation}\label{eq:hamineq}
    H(x^*(t),u^*(t),p^*(t),p_0^*) 
    \geq 
    H(x^*(t),u,p^*(t),p_0^*),
\end{equation}
with $x^*:[t_0,t_f]\rightarrow \mathbb{R}^2$ being the optimal trajectory associated with the control $u^*$. From the PMP, recall that the costates \(p^* = (p_1^*, p_2^*):[t_0,t_f]\rightarrow \mathbb{R}^2\) exist and satisfy 
\(\dot{p}^* = -\left. H_x \right|_{*}\). 
Hence, the solutions of the costate equations for the double integrator are
\begin{equation}\label{eq:soldicostat}
    p_1^*(t) = c_1, \quad 
    p_2^*(t) = c_2 - c_1 (t - t_0),
\end{equation}
where \(p_1^*(t_0) = c_1\) and \(p_2^*(t_0) = c_2\).
Since the Hamiltonian~\eqref{eq:hamDI} is linear in the control \(u\), 
Eq.~\eqref{eq:hamineq} reduces to
\begin{equation}\label{eq:ineqDI}p_2^*(t)\,u^* \geq p_2^*(t)\,u,\end{equation} 
which must hold for all \(t \in [t_0, t_f]\) and for all \(u \in [u_{\min}, u_{\max}]\).
From Eq.~\eqref{eq:soldicostat}, \(p_2^*(t)\) is linear (affine) and thus changes sign at most once. The optimal control is therefore bang-bang, taking the value \(u_{\max}\) when \(p_2^*(t) > 0\) 
and \(u_{\min}\) when \(p_2^*(t) < 0\), with at most one switching time.

In the following, we carry out an analogous reasoning concerning what we have called the relativistic double integrator given by Eq.~\eqref{eq:reldi}. Starting from its Hamiltonian
\begin{equation}\label{eq:Hamwrap}
    %H(x,u,p,p_0) = p_0 + p_1 x_2 + p_2 u \left(1 - \frac{x_2^2}{c^2}\right)^{3/2},
    H(x,u,p,p_0) = p_0 + p_1 x_2 + p_2 u \left(1 - (x_2/c)^2\right)^{3/2},
\end{equation}
we observe that the Hamiltonian of the relativistic double integrator corresponds to that of the standard double integrator~\eqref{eq:hamDI}, with an additional term that becomes negligible when $|\dot q| \ll c$.
Recalling that $\dot{p}^* = -\left. H_x \right|_{*}$, the costate dynamics are given by 
\begin{equation}
    \begin{cases}
        \dot p_1^* = -\left. H_{x_1}\right|_{*}= 0\\
        \dot p_2^* = -\left. H_{x_2}\right|_{*} = -p^*_1 +3p^*_2x^*_2\frac{u^*}{c^2}\sqrt{1-\frac{x_2^{*2}}{c^2}}\\
        %\dot p_2^* = -\left. H_{x_2}\right|_{*} = -p^*_1 +3x^*_2\frac{u^*}{c^2}\sqrt{1- (x_2^*/c)^2}\\
        p_1^*(t_0) = c_1, \quad p_2^*(t_0)=c_2
    \end{cases}.
\end{equation}
Solving the costate equations, we obtain
\begin{equation}\label{eq:relcostsol}
    p^*_1(t) = c_1, \qquad p^*_2(t) = \mu(u^*,t)^3\zeta(t),
\end{equation}
where the factor $\zeta$ is defined as
    \begin{equation}\label{eq:zeta}\zeta(t)=\frac{c_2}{\gamma_0^3} - \frac{c_1}{u^*}(x_2^*(t)-\dot q_0),\end{equation}
and $\mu$ and $x_2^*$ as in Eq.~\eqref{eq:gammu} and \eqref{eq:qdot}, respectively.

Following the PMP, if an optimal control $u^*$ exists, it maximizes the Hamiltonian pointwise; hence, by writing Eq.~\eqref{eq:hamineq} explicitly, the analog of Eq.~\eqref{eq:ineqDI} becomes
\begin{equation}\label{eq:ineqDI}p_2^*(t)\,u^*\,\,\left(1-\frac{x_2^{*2}}{c^2}\right)^{3/2}\geq p_2^*(t)\,u\,\left(1-\frac{x_2^{*2}}{c^2}\right)^{3/2},\end{equation} 
which should hold for all \(u \in U\) and for all \(t \in [t_0,t_f]\). Note that a straightforward computation shows that %$\left(1-(x_2^*/c)^2\right)^{3/2}=\mu(u,t)^{-3}$,
\begin{equation}
    \left(1-(x_2^*/c)^2\right)^{3/2}=\mu(u,t)^{-3},
\end{equation}for all $t\in[t_0,t_f]$ and for all $u\in U$. Hence, Eq.~\eqref{eq:ineqDI} becomes 
\begin{equation}\label{eq:ineq}\zeta(t)u^*\geq \zeta(t)\,u.\end{equation}
Therefore, the possible bang-bang control depends on the sign of $\zeta(t)$ as follows \begin{equation}
u^*(t) =
\begin{cases}
u_{\max}, &  \text{if } \zeta(t) > 0\\
u_{\min}, & \text{if } \zeta(t) < 0\
\end{cases}.
\end{equation}
The derivative of $\zeta$ is given by $\dot \zeta(t)= -\frac{c_1}{\mu^3}$, and, since $\mu(u,t)>0$ for all \(u \in U\) and for all \(t \in [t_0,t_f]\), it can be concluded that $\zeta$ is monotonic. Therefore, it intersects the time axis at most one time (depending on $c_1,c_2$), leading to a time-optimal control of bang-bang type with at most one switching time as in the classical double integrator. 

To argue for the existence of optimal controls we show that for a given time interval $[t_0,t_f]$ 
the set of reachable states within time $t$, denoted by $R^t(x(t_0))$, is compact for any $t\in[t_0,t_f]$, from which existence follows~\cite[Ch.~4]{Liberzon:12}. Because $U$ is compact, for any $x=(x_1,x_2)$ with $|x_2|<c$, the set $\left\{f(x,u)=\left[x_2,(1-x_2^2/c^2)^{3/2}u\right]^T : u\in U \right\}$ is a line segment in the $\dot{x_1}, \dot{x}_2$ plane, hence it is compact and convex. Then, by Filippov's theorem, $R^t(x(t_0))$ is compact for each $t\in[t_0,t_f]$.

\section{Limiting the velocity using special relativity}\label{sec:nonlin}
This section presents an application of the relativistic double integrator introduced in Section \ref{sec:rel}. Specifically, we propose to limit the speed of a system governed by double integrator dynamics, that is, ensure that $|\dot q(t)| < \vmax$ for an arbitrary $\vmax >0$ for all $t>t_0$, using the explicit relationship given by Eq.~\eqref{eq:reldi}. To this end, we substitute $c$ with $\vmax$. We refer to this procedure of modifying the input as the \emph{Minkowski wrap}, and we call the resulting system the \emph{wrapped system}. In this section, we introduce the idea of Minkowski wrap and show its application to wrapping a double integrator with inputs generated by proportional and time-optimal control laws.

\subsection{Minkowski Wrap}
To simplify the notation, we first write the double integrator system in state-space form 
\begin{equation}\label{eq:double_integ_ss}
\begin{aligned}
    \dot{x} & =\begin{bmatrix}
    \dot x_1 & \dot x_2
\end{bmatrix}^T
\end{aligned}
\end{equation}
in which $x:= \begin{bmatrix}
    q & \dot q
\end{bmatrix}^T$, $A=\begin{bmatrix}
        0 & 1 \\ 0 & 0
    \end{bmatrix}$, and $B=\begin{bmatrix}
        0 \\ 1
    \end{bmatrix}$. 

Suppose it is desired that the trajectories of the double integrator system given above satisfy that $|x_2| < v_{\text{max}} $.  
Using the previously introduced relativistic notions, we introduce a mapping $\sigma : U \times X\rightarrow U$ defined by 
\begin{equation}
\sigma (u,x)=u\left(1-\frac{x_2^2}{v_{\text{max}}^2} \right)^{3/2}.
\end{equation}
Substituting the input $u$ with $\psi=\sigma (u,x)$ in double integrator system yields the relativistic double integrator 
\begin{equation}
\begin{aligned}
 \dot x_1 &= x_2 \\
 \dot x_2 &= \psi=u\left(1-\frac{x_2^2}{v_{\text{max}}^2} \right)^{3/2}
\end{aligned}
\end{equation}
with the exception that in the place of speed of light $c$ we use an arbitrary speed limit $v_{\text{max}}$. Therefore, the trajectories inherently satisfy the limit on $x_2$, that is, they satisfy that $|x_2|<v_{\text{max}}$ for all $t>t_0$ (see Remark~\ref{rem:bounded_vel}). 

\begin{figure}
    \centering
    \includegraphics[width=0.9\columnwidth]{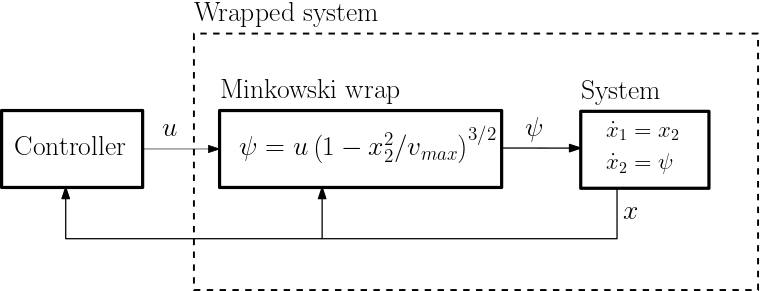}
    \caption{Diagram illustrating the Minkowski wrap.}
    \label{fig:MinkDiag}
    \vspace{-1.15em}
\end{figure}

Modifying the input to satisfy Eq.~\eqref{eq:reldi} requires feedback and induces a non-linearity into the system dynamics with respect to the input. Therefore, in manner analogous to feedback linearization (see \cite[Ch.~13]{Khalil:14}), this procedure can be seen as a form of \emph{feedback non-linearization}. Fig.~\ref{fig:MinkDiag} shows a diagram illustrating this idea.

The intuitive idea is that, in the Minkowski law \eqref{eq:dyn}, or equivalently in the relativistic double integrator \eqref{eq:reldi}, the Lorentz factor $\gamma$ in Eq.~\eqref{eq:gam} contributes to slowing down a particle that, under constant acceleration, approaches the speed limit. Clearly, this is not the only way to slow down a system approaching a velocity bound; it is simply the direct adaptation of the approach from special relativity. Namely, one could consider a simpler factor, such as $\gamma' = 1 - |\dot q|/v_{\max}$, but this would not be differentiable everywhere. In this sense, the Lorentz factor can be seen as the simplest smooth option to slow down a system.%particle.

\subsection{Linear state-feedback control}
Given the double integrator system \eqref{eq:double_integ_ss}, let $u(t)=-Kx(t)$ with $K=[k_1\; k_2]^T$ be a state-feedback control law that stabilizes the system about the origin $[0,0]^T$. Whereas the control law ensures stability, it does not (necessarily) satisfy any bounds on the resulting state trajectories. 

The wrapped system in the state space form is expressed by the following equations
\begin{equation}\label{eq:ss_wrapped_syss}
\begin{aligned}
\dot{x}_1 &= x_2 \\   
\dot{x}_2 &= (-k_1x_1 - k_2x_2)\left(1-(x_2/v_{\text{max}})^2\right)^{3/2}.
\end{aligned}
\end{equation}
To show that the origin of the double integrator system is still stable, we rely on Lyapunov stability analysis. Specifically, we show that there exists a Lyapunov function $V(x)$, for the wrapped system given in \eqref{eq:ss_wrapped_syss}. We show this by construction. Let a Lyapunov function candidate be
\begin{equation}\label{eq:Lyap_cand}
    V(x)=V_1(x_1) + V_2(x_2).
\end{equation}
Similar to an energy function, we select $V_1(x_1)=\frac{1}{2}k_1x_1^2$ and select $V_2(x_2)$ so that it satisfies
\begin{equation}\label{eq:dV2}
\frac{dV_2(x_2)}{dx_2} = \frac{x_2}{\left(1-{x_2^2}/{v_{\text{max}}^2} \right)^{3/2}}.
\end{equation}
Integrating \eqref{eq:dV2} results in 
\begin{equation}
    V_2(x_2) = \dfrac{v_{\text{max}}^2}{\sqrt{1-{x_2^2}/{v_{\text{max}}^2}}} + C,
\end{equation}
in which $C$ is a constant. Imposing $V_2(0)=0$ yields 
\begin{equation}
    V_2(x_2)=v_{\text{max}}^2\left( \frac{1}{\sqrt{1-{x_2^2}/{v_{\text{max}}^2}}} -1\right).
\end{equation}
Substituting \(V_1\) and \(V_2\) into \eqref{eq:Lyap_cand} with their respective expressions, we obtain
\begin{equation}
    V(x)=\frac{1}{2}k_1x_1^2 + v_{\text{max}}^2\left( \frac{1}{\sqrt{1-{x_2^2}/{v_{\text{max}}^2}}} -1\right).
\end{equation}
Noting that $k_1,k_2>0$ for the control law to be stabilizing, $V(x)$ is positive definite in domain $D:=\{ x \in \R^2 \mid |x_2|< c \}$. The derivative of $V(x)$ along trajectories of the system is given by 
\begin{equation}\label{eq:dV}
\begin{aligned}
\dot{V}(x) & :=\frac{\partial V} {\partial x_1} \dot x_1 + \frac{\partial V} {\partial x_2} \dot x_2 \\
&= k_1 x_1 x_2 + \frac{x_2(-k_1x_1 - k_2x_2)}{\left(1-{x_2^2}/{v_{\text{max}}^2} \right)^{3/2}} \left(1-{x_2^2}/{v_{\text{max}}^2} \right)^{3/2} \\
&=-k_2x_2^2.
\end{aligned}
\end{equation}
It follows that $\dot{V}(x)<0$ for all $x \in D, x\neq 0$ since $k_2>0$. From which we conclude that the origin of the wrapped closed-loop system in \eqref{eq:ss_wrapped_syss} is locally asymptotically stable. Note that the result naturally extends stabilizing around any value for $x_1$ through a change of variables.  

\begin{figure}
    \centering
    \includegraphics[width=0.8\linewidth]{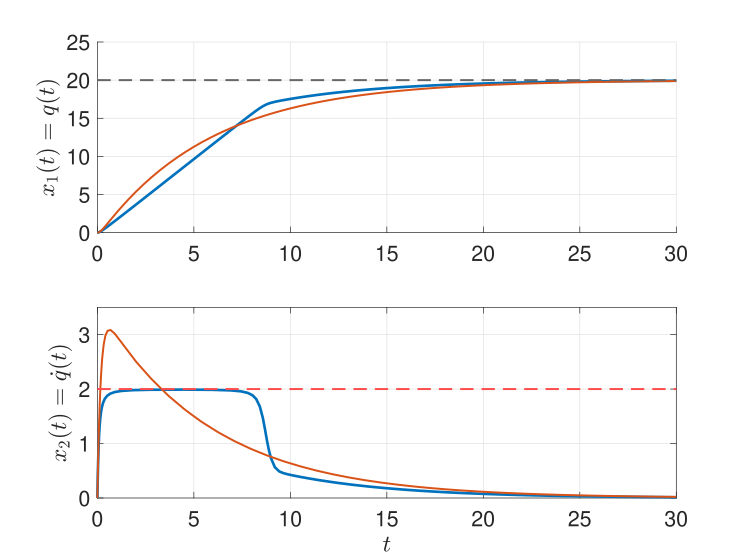}
    \caption{Comparing state-feedback control for stabilizing a double integrator system at $x^\circ=[20,\;0]^T$ with and without Minkowski wrap. Orange lines refer to state trajectories for double integrator without Minkowski wrap and blue are with Minkowski wrap. \textit{Up}: The dashed line indicates the reference. \textit{Down}: Red dashed line indicates the limit on $x_2$.}
    \label{fig:MinkWrapLinFB}
    \vspace{-1em}
\end{figure}

Fig.~\ref{fig:MinkWrapLinFB} presents a simple illustrative example. In the example problem, it is desired to steer the state to a reference given by $x^\circ = [20, \;0]^T$ while satisfying that $|x_2(t)| < 2$ for all $t>0$. With a change of variables $z = x-x^\circ$, a new double integrator system is obtained as $\dot{z} = (A-BK)z$, in which $A$ and $B$ are from Eq.~\eqref{eq:double_integ_ss} and $K= [k_1,\;k_2]=[1,\; 6]$ which renders the closed-loop system stable about the origin. Consecutively, the wrapped system is determined through Eq.~\eqref{eq:ss_wrapped_syss}. Comparing the state trajectories with and without Minkowski wrap, in both cases the system stabilizes at $x^\circ$. Without Minkowski wrap, the limit on $x_2$ is violated. 

Although we did not consider any bounds on the control, adding saturation to the control inputs would not change the results. Stability under control saturation has been addressed for state-feedback control of linear systems, under general linear feedback laws~\cite{kapasouris1988design}, or for linear quadratic regulator type of problems~\cite{Gokcek:01}. Therefore, as long as the control law is stabilizing, the speed limits can be imposed using the Minkowski wrap. Compared to approaches based on the joint design of CLF-CBF (e.g., \cite{Ames:14}), our method is based on an existing stabilizing controller and imposes velocity constraints without the need to solve an optimization problem. 

\subsection{Time-optimal control}
\begin{figure}
    \centering
    \includegraphics[width=1\linewidth]{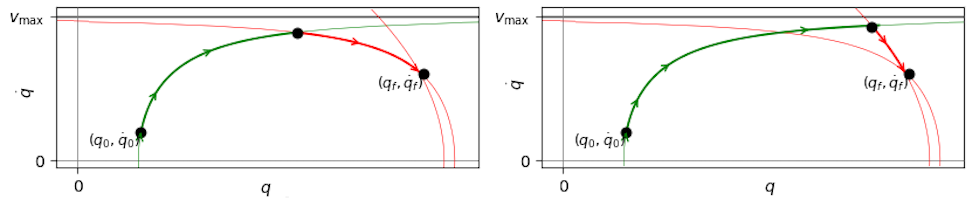} 
    \caption{\textit{Left:} A bang-bang trajectory of the relativistic double integrator in the upper half-plane with a single switch. \textit{Right:} A bang-bang trajectory for the same initial and final conditions, where trajectories of the relativistic double integrator are used when approaching the velocity bound, and standard double-integrator trajectories are used when moving away from it, both double integrators have the same control interval $U=[-u_{\max}, u_{\max}]$.}
    \label{fig:squish}
    \vspace{-0.2em}
\end{figure}

\begin{figure}
    \centering
    \includegraphics[width=0.7\columnwidth]
    {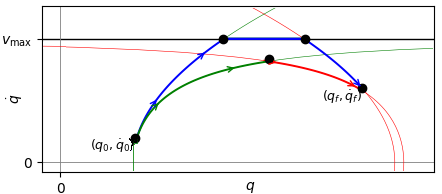} 
    \caption{Comparison between a bang-bang trajectory of the standard double integrator (in blue), clipped when the maximum speed is reached, and a bang-bang trajectory of the relativistic double integrator (in green and red), both having the same control interval, initial and final conditions.}
    \label{fig:clip}
    \vspace{-1em}
\end{figure}

In Fig.~\ref{fig:squish} (\textit{Left}), we give an illustrative example of the time-optimal solution characterized in Section \ref{sec:pontryagin}, with one single switch. We consider a system with dynamics described by Eq.~\eqref{eq:reldi}, and initial and final configurations $(q_0, \dot{q}_0)$ and $(q_f, \dot{q}_f) \in \mathbb{R}^2$. 
An immediate comment regarding Fig.~\ref{fig:squish} (\textit{Left}) %fig:ex}
is that the control, although time-optimal for the relativistic double integrator, appears slow, for example, when compared to the parabolic trajectories of the standard double integrator (first column in Fig.~\ref{fig:all_phases}) that are clipped once the maximum speed is reached, see Fig.~\ref{fig:clip}. In fact, the clipped bang-bang solution is time-optimal for the double integrator, and thus remains time-optimal when a velocity bound is added. However, notice that clipping the bang-bang solutions in the phase portrait of the double integrator introduces an additional singularity, unlike what happens for the relativistic double integrator.

A partial solution is to propose a hybrid between relativistic and non-relativistic double integrators. The idea is that a Minkowski wrap is needed only when the system is dangerously directed toward the speed bound, for instance when following a green trajectory in the upper half-plane or a red trajectory in the lower half-plane (Fig.~\ref{fig:all_phases}, second column). Whereas, when a system moves away from the speed limit, towards rest (red trajectories in the upper-half plane and green in the lower half), one could use a standard double integrator to save time, as depicted in Fig.~\ref{fig:squish}. We must stress that, given initial and final conditions as in Fig.~\ref{fig:squish}, the time at which the switch occurs is unique, in both Fig.~\ref{fig:squish} (\textit{Left}) and Fig.~\ref{fig:squish} (\textit{Right}). In fact, since there exists a unique green curve passing through $(q_0,\dot q_0)$ and a unique red curve passing through $(q_f,\dot q_f)$, the switching time is determined by their unique intersection point in the upper half-plane.%the red and green curves are obtained with the fixed values $u$

\section{Discussion and Conclusions}
In this work, we have analyzed a relativistic version of the double integrator system, which arises as a consequence of the Minkowski law, or relativistic form of Newton's law. A key result is that the time-optimal control for a relativistic double integrator is bang-bang and, by application of the PMP, is shown to have at most one switch. The speed of light $c$ appearing in the relativistic double integrator dynamics can be replaced by an arbitrary velocity bound, effectively limiting the attainable velocity along the system's trajectories. Taking advantage of this observation, we introduced the notion of \emph{Minkowski wrap} which reshapes the input to a double integrator system resulting in a non-linear system with relativistic double integrator dynamics, intrinsically satisfying the velocity bounds without clipping. 

This notion is particularly relevant for safety critical systems. It can be seen as a user pressing the accelerator, which gradually becomes increasingly ineffective for his/her own safety, as opposed to abruptly setting acceleration to zero once the speed limit is reached. 

Future work will extend this approach to higher-order systems and to  
systems with general proper accelerations expressed in inertial coordinates, with the aim of deriving analytical (as for the double integrator) or numerical solutions, and further investigations of time-optimal controls. Higher-dimensional systems composed of a vector of double integrators were considered in prior work~\cite{LaValle:23} and it was shown that a class of time-optimal solutions for boundary conditions that are not rest-to-rest require time-stretching, that is, maximum of the minimum times does not work. Extending this result to the case of relativistic double integrator is also an interesting future direction.

The results on characterizing the time-optimal controls naturally extend to linear systems that can be transformed into a double integrator by change of coordinates, if the constraints are meaningful in the new coordinates. We expect that the ideas can potentially be extended to a more general class of nonlinear stabilizable second-order systems of the form $\ddot q = f(q, \dot q, u)$, in which $f$ is global Lipschitz continuous in all its arguments. In this case, one can describe the system as a double integrator $\ddot q = \psi$, in which $\psi \in \Psi(q,\dot q)$ is the acceleration and $\Psi(q, \dot q)$ is a state-dependent set of admissible accelerations (see \cite{LaValle:23} for an example). Suppose there exists a set $\bar{\Psi}$ satisfying that $\bar{\Psi} \supseteq \Psi(q, \dot q)$ for all $q, \dot q$. Then, the system $\ddot q = f(q, \dot q, u)$ is ``dominated" by the double integrator $\ddot q = \psi$, in which $\psi \in \bar{\Psi}$. Therefore, the Minkowski wrap of the dominating double integrator can be applied to ``wrap" the trajectories of the system. Exploring this direction is left for future work.

%%%%%%%%%%%%%%%%%%%%%%%%%%%%%%%%%%%%%%%%%%%%%%%%%%%%%%%%%%%%%%%%%%%%%%%%%%%%%%%%

\vspace{-0.5 cm}
\bibliographystyle{IEEEtran}
\bibliography{citations}
\end{document}

%% file: defs.tex
\def\vmax{v_{\text{max}}}